\documentclass[12pt]{article}
\usepackage{amsfonts}
\usepackage{amssymb}
\usepackage{graphics,amsmath}
\usepackage{physics}
\usepackage{textcomp}
\usepackage{cite}

\def\hybrid{\topmargin -20pt    \oddsidemargin 0pt
        \headheight 0pt \headsep 0pt
        \textwidth 6.35in       % BS paper
        \textheight 9.25in       % BS paper
        \marginparwidth .875in
        \parskip 5pt plus 1pt   \jot = 1.5ex}

\hybrid

\def\baselinestretch{1.2}

\catcode`\@=11

\def\marginnote#1{}
\newcount\hour
\newcount\minute
\newtoks\amorpm
\hour=\time\divide\hour by60
\minute=\time{\multiply\hour by60 \global\advance\minute by-\hour}
\edef\standardtime{{\ifnum\hour<12 \global\amorpm={am}%
        \else\global\amorpm={pm}\advance\hour by-12 \fi
        \ifnum\hour=0 \hour=12 \fi
        \number\hour:\ifnum\minute<10 0\fi\number\minute\the\amorpm}}
\edef\militarytime{\number\hour:\ifnum\minute<10 0\fi\number\minute}
\def\draftlabel#1{{\@bsphack\if@filesw {\let\thepage\relax
   \xdef\@gtempa{\write\@auxout{\string
      \newlabel{#1}{{\@currentlabel}{\thepage}}}}}\@gtempa
   \if@nobreak \ifvmode\nobreak\fi\fi\fi\@esphack}
        \gdef\@eqnlabel{#1}}
\def\@eqnlabel{}
\def\@vacuum{}
\def\draftmarginnote#1{\marginpar{\raggedright\scriptsize\tt#1}}

\def\draft{\oddsidemargin -.5truein
        \def\@oddfoot{\sl preliminary draft \hfil
        \rm\thepage\hfil\sl\today\quad\militarytime}
        \let\@evenfoot\@oddfoot \overfullrule 3pt
        \let\label=\draftlabel
        \let\marginnote=\draftmarginnote
   \def\@eqnnum{(\theequation)\rlap{\kern\marginparsep\tt\@eqnlabel}%
\global\let\@eqnlabel\@vacuum}  }

\def\preprint{\twocolumn\sloppy\flushbottom\parindent 2em
        \leftmargini 2em\leftmarginv .5em\leftmarginvi .5em
        \oddsidemargin -.5in    \evensidemargin -.5in
        \columnsep .4in \footheight 0pt
        \textwidth 10.in        \topmargin  -.4in
        \headheight 12pt \topskip .4in
        \textheight 6.9in \footskip 0pt
        \def\@oddhead{\thepage\hfil\addtocounter{page}{1}\thepage}
        \let\@evenhead\@oddhead \def\@oddfoot{} \def\@evenfoot{} }

\def\numberbysection{\@addtoreset{equation}{section}
        \def\theequation{\thesection.\arabic{equation}}}

\def\underline#1{\relax\ifmmode\@@underline#1\else
        $\@@underline{\hbox{#1}}$\relax\fi}

\def\titlepage{\@restonecolfalse\if@twocolumn\@restonecoltrue\onecolumn
     \else \newpage \fi \thispagestyle{empty}\c@page\z@
        \def\thefootnote{\fnsymbol{footnote}} }

\def\endtitlepage{\if@restonecol\twocolumn \else \newpage \fi
        \def\thefootnote{\arabic{footnote}}
        \setcounter{footnote}{0}}  %\c@footnote\z@ }

\catcode`@=12
\relax

\def\figcap{\section*{Figure Captions\markboth
        {FIGURECAPTIONS}{FIGURECAPTIONS}}\list
        {Figure \arabic{enumi}:\hfill}{\settowidth\labelwidth{Figure
999:}
        \leftmargin\labelwidth
        \advance\leftmargin\labelsep\usecounter{enumi}}}
 \relax
\def\tablecap{\section*{Table Captions\markboth
        {TABLECAPTIONS}{TABLECAPTIONS}}\list
        {Table \arabic{enumi}:\hfill}{\settowidth\labelwidth{Table
999:}
        \leftmargin\labelwidth
        \advance\leftmargin\labelsep\usecounter{enumi}}}
 \relax
\def\reflist{\section*{References\markboth
        {REFLIST}{REFLIST}}\list
        {[\arabic{enumi}]\hfill}{\settowidth\labelwidth{[999]}
        \leftmargin\labelwidth
        \advance\leftmargin\labelsep\usecounter{enumi}}}
 \relax
\makeatletter
\newcounter{pubctr}
\def\publist{\@ifnextchar[{\@publist}{\@@publist}}
\def\@publist[#1]{\list
        {[\arabic{pubctr}]\hfill}{\settowidth\labelwidth{[999]}
        \leftmargin\labelwidth
        \advance\leftmargin\labelsep
        \@nmbrlisttrue\def\@listctr{pubctr}
        \setcounter{pubctr}{#1}\addtocounter{pubctr}{-1}}}
\def\@@publist{\list
        {[\arabic{pubctr}]\hfill}{\settowidth\labelwidth{[999]}
        \leftmargin\labelwidth
        \advance\leftmargin\labelsep
        \@nmbrlisttrue\def\@listctr{pubctr}}}
 \relax
\makeatother
\newskip\humongous \humongous=0pt plus 1000pt minus 1000pt

\newif\ifdtup

\relax

\def\be{\begin{equation}}
\def\ee{\end{equation}}
\def\ba{\begin{eqnarray}}
\def\ea{\end{eqnarray}}

\def\no{\noindent}

\def\IR{\relax{\rm I\kern-.18em R}}
\def\II{\relax{\rm 1\kern-.35em1}}

\renewcommand{\theequation}{\thesection.\arabic{equation}}
\csname @addtoreset\endcsname{equation}{section}

\def\IR{\relax{\rm I\kern-.18em R}}
\def\inv{^{\raise.15ex\hbox{${\scriptscriptstyle -}$}\kern-.05em 1}}

\def\dif{{\textnormal d}}

\DeclareMathOperator{\cosech}{cosech}
\DeclareMathOperator{\D}{D}
\DeclareMathOperator{\cotanh}{cotanh}
\DeclareMathOperator{\arctanh}{arctanh}

\DeclareMathOperator{\B}{B}
\DeclareMathOperator{\e}{e}
\DeclareMathOperator{\Ei}{Ei}
\DeclareMathOperator{\AdS}{AdS}
\DeclareMathOperator{\Esfera}{S}
\DeclareMathOperator{\Toro}{T}

\begin{document}

\begin{titlepage}
\begin{center}

\vskip .5in

{\LARGE Quantum corrections to minimal surfaces with mixed three-form flux}

\vskip 0.4in

{\bf Rafael Hern\'andez$^1$},  \phantom{x} {\bf Juan Miguel Nieto$^2$} \phantom{x} and \phantom{x} {\bf Roberto Ruiz$^1$} 
\vskip 0.1in

${}^1\!$
Departamento de F\'{\i}sica Te\'orica \\ and Instituto de F\'{\i}sica de Part\'{\i}culas y del Cosmos, IPARCOS \\
Universidad Complutense de Madrid \\
$28040$ Madrid, Spain \\

\vskip .2in

${}^2\!$
Department of Mathematics, University of Surrey \\ Guildford, GU2 7XH, UK \\

\vskip .2in

{\footnotesize{\tt rafael.hernandez@fis.ucm.es, j.nietogarcia@surrey.ac.uk, roruiz@ucm.es}}

\end{center}

\vskip .4in

\centerline{\bf Abstract}
\vskip .1in
\no
\noindent
We obtain the ratio of semiclassical partition functions for the extension under mixed flux of the minimal surfaces subtending a circumference 
and a line in Euclidean $\AdS_{3}\times\Esfera^{3}\times\Toro^{4}$. We reduce the problem to the computation of a set of functional determinants. 
If the Ramond-Ramond flux does not vanish, we find that the contribution of the $B$-field is comprised in the conformal anomaly. 
In this case, we successively apply the Gel'fand-Yaglom method and the Abel-Plana formula to the flat-measure determinants. To cancel the resultant infrared divergences, 
we shift the regularization of the sum over half-integers depending on whether it corresponds to massive or massless fermionic modes. We show that the result 
is compatible with the zeta-function regularization approach. In the limit of pure Neveu-Schwarz-Neveu-Schwarz flux we argue that the computation trivializes. 
We extend the reasoning to other surfaces with the same behavior in this regime. 
\vskip .4in
\noindent

\end{titlepage}

\vfill
\eject

\def\baselinestretch{1.2}

%%%%%%%%%%%%%%%%%%%%%%%%%%%%%%%%%%%%%%%%%%%%%%%%%%%%%%%%%%%%%%%%%%%%%%%%

\baselineskip 20pt

%%%%%%%%%%%%%%%%%%%%%%%%%%%%%%%%%%%%%%%%%%%%%%%%%%%%%%%%%%%%%%%%%%%%%%%%
%%%%%%%%%%%%%%%%%%%%%%%%%%%%%%%%%%%%%%%%%%%%%%%%%%%%%%%%%%%%%%%%%%%%%%%%

\section{Introduction}

The connection between Wilson loops and minimal surfaces raised a milestone in the AdS/CFT correspondence. 
The original proposal establishes that the strong coupling limit of the expectation value of a Wilson loop in $\mathcal{N}=4$ supersymmetric Yang-Mills  
in four dimensions is given by the regularized minimal area swept by a string probe propagating in Euclidean $\AdS_{5}\times\Esfera^{5}$ and terminating in the contour of the Wilson loop on the boundary 
of the Euclidean anti-de Sitter space~\cite{Maldacena, Rey}. The approach to the evaluation of Wilson loops at strong coupling triggered a profusion of developments, 
to which the link between minimal surfaces and the classical integrable structure underlying the AdS/CFT correspondence belongs. 
In reference~\cite{loiF} it was shown that the periodic ansatz for spinning strings employed in~\cite{NR} may be extended to the study of world-sheets with open boundary conditions 
at the boundary of the Euclidean anti-de Sitter space. The use of that kind of ansatz allowed the reduction of the problem of finding minimal area surfaces in Euclidean $\AdS_{5}\times\Esfera^{5}$ 
to the construction of solutions of the Neumann-Rosochatius integrable system. Since one can systematically obtain the latter in terms of elliptic and hyperelliptic functions, it was then possible to cover the string picture 
for a wide range of Wilson loop configurations. A complementary series of developments concerned the one-loop quantization of these minimal surfaces, which led to the comparison between both sides of the duality 
beyond the leading order.  In reference~\cite{Drukker}, the one-loop effective action of various minimal surfaces was expressed through the ratio of functional determinants of certain differential operators. 
The ratio of the semiclassical partition functions of the surfaces subtending a circle and a line at the boundary presented in~\cite{Drukker} was explicitly obtained in reference~\cite{Kruczenski}. 
The connection between integrability and Wilson loops was shown to emerge in this context through the effective background for the fluctuations, where the solution to the integrable mechanical system comes again into play. 
This setting permitted the decomposition of every two-dimensional determinant into the product of one-dimensional determinants by means of the boundary conditions of the fields with respect 
to the Euclidean time world-sheet coordinate. The resultant one-dimensional determinants were evaluated using the Gel'fand-Yaglom method, which renders the derivation 
of the eigenvalues of the differential operators and the evaluation of their product into the resolution of an initial value problem. The product over one-dimensional determinants 
was then performed in the zeta-function regularization scheme and it was shown to be in agreement with the gauge theory, up to a normalization factor which was later retrieved in~\cite{Factor}. 
Such an approach to the quantization of minimal surfaces has paved the way of several complementary lines 
of research~\cite{Factor,Hou,Chu,Faraggi,Buchbinder,Kim,Forini,Kristjansen,Dekel,Griguolo,Giangreco,Pando,Zarembo,Beccaria,Cagnazzo,Silva,Aguilera,Daniel,David}.

The relevance of minimal area surfaces in the AdS$_{5}$/CFT$_{4}$ correspondence has motivated their study in lower-dimensional avatars of the duality. One of the contributions along these lines, 
within the framework of type IIB string theory on $\AdS_{3}\times\Esfera^{3}\times\Toro^{4}$, was the construction in~\cite{Minimal} of minimal surfaces in Euclidean $\AdS_{3}$ with mixed Ramond-Ramond (R-R) 
and Neveu-Schwarz-Neveu-Schwarz (NS-NS) three-form fluxes on the basis of the underlying integrable mechanical model (see~\cite{DavidSad} for other findings of minimal surfaces under the presence of mixed fluxes). 
The authors employed the periodic ansatz of~\cite{loiF} to study the extension of the class of classical world-sheets subtending two concentric circumferences at the boundary of 
Euclidean anti-de Sitter space by the introduction of NS-NS flux. They found that the NS-NS flux either brings the world-sheet near to the boundary or separates the circumferences of the annulus at the boundary. 
When the R-R flux vanishes the classical world-sheet adheres to the boundary in the former case, whereas the outer radius diverges in the latter case. 
The surface that subtends a circle plays the role of a threshold solution, in the sense that in the limit of pure NS-NS flux the world-sheet adheres to the boundary and it ends in 
an annulus whose outer radius diverges.~\footnote{It was also proved in reference~\cite{Minimal} that the algebraic elliptic curve that describes the solutions becomes singular in the regime 
of pure NS-NS flux.} In view of these distinctive features, it is natural to pose the question of the behavior of these minimal surfaces when quantum corrections are taken into account. 
In this article we will analyze the one-loop effective action of the extension under fluxes of the minimal area surface subtending a circumference. This solution constitutes an appropriate framework 
for the study of the mixed flux regime in the semiclassical picture, since it is simple enough to allow a tractable analysis but still comprises the
major features that are meant to be brought in. In order to proceed, we will follow the analysis of~\cite{Kruczenski} and introduce the deformation under mixed fluxes of the classical world-sheet 
subtending a line as a reference solution, as it shares the same behavior with this surface in the vicinity of the boundary. In this way, we will be able to consider the ratio of both semiclassical partition functions, 
for which infrared divergences are expected to cancel.

The remaining part of the article is structured as follows. In section~\ref{Solutions} we will present the two classical solutions whose one-loop effective action is going to be computed. 
In section~\ref{soD} we will employ the background field expansion around these configurations to find the set of relevant functional determinants in both cases. 
We will start from the statically gauge-fixed Nambu-Goto action together with the quadratic truncation of the fermionic Lagrangian, and argue that the ghost determinant is compensated by field redefinitions 
of the quadratic fluctuations. We will show that the contribution of the NS-NS term is enclosed in the conformal anomaly in the mixed flux regime, and hence the computation of determinants 
with flat measure remains unaffected by the presence of the $B$-field. On the contrary, the limit of pure NS-NS will require a separated analysis. In section~\ref{serT} we will consider the flat-measure determinants 
arising in the pure R-R flux regime. We will resort to the Gel'fand-Yaglom method and the Abel-Plana formula to show that infrared divergences cancel in the ratio of the semiclassical partition functions. 
We will then compute the difference between the corresponding one-loop effective actions. This problem will lead us to introduce a shift in the regularization of the sum over half-integers 
massless fermionic modes into a sum over integers  with respect to the massive case. We will compare the expected result with the answer given by the zeta-function regularization prescription and show that both agree. 
In section~\ref{Neveu} we will discuss the pure NS-NS flux limit of our solution. We will show that the problem trivializes when the classical world-sheet adheres to the boundary. 
We will also argue that a similar phenomenon extends to the remaining surfaces in the class of solutions of~\cite{Minimal} that remain stuck at the boundary in this limit, even if the factorization 
of the NS-NS flux into the conformal anomaly breaks down. In section~\ref{Conclusions} we will summarize our results and comment on some possible future developments. 
We have relegated the details on the application of the Gel'fand-Yaglom method and the Abel-Plana formula to the appendices. 

%%%%%%%%%%%%%%%%%%%%%%%%%%%%%%%%%%%%%%%%%%%%%%%%%%%%%%%%%%%%%%%%%%
%%%%%%%%%%%%%%%%%%%%%%%%%%%%%%%%%%%%%%%%%%%%%%%%%%%%%%%%%%%%%%%%%%

\section{Classical solutions} 

\label{Solutions}

In this section, we will present the classical solutions whose semiclassical partition function will be computed below. The first solution that we will address subtends a circumference
on the boundary of Euclidean three-dimensional anti-de Sitter space which splits into two concentric circles when the NS-NS flux is introduced. It has winding index $k=1$ along both circumferences, 
and zero classical dilatation charge. When the R-R flux vanishes, the ratio of the two radii either diverges or goes to zero, and the classical world-sheet completely adheres to the boundary. 
We will then move to the surface that subtends a strip at the boundary in the presence of NS-NS flux. This surface will play the role of reference solution. This strip shrinks to a line in the pure R-R flux limit. 
On the contrary, the distance between both lines diverges in the limit of pure NS-NS flux, where the surface again remains stuck at the boundary. 

\subsection{Surface subtending two concentric circumferences}

This classical solution is conveniently expressed in the parameterization of the Poincar\'e patch of Euclidean AdS$_{3}$ through the coordinates
\be
u = \frac{z}{r} \ , \quad v=\log\sqrt{z^2+r^2} \ ,
\ee
together with the polar angle $\theta$ of the plane at the conformal boundary. Accordingly, the metric reads
\be
\label{Metric}
\dif s^2 = \frac{1}{u^2}\left[\dif \theta^2+\frac{\dif u^2}{1+u^2}+\left(1+u^2\right)\dif v^2\right] \ ,
\ee
with the conformal boundary at $u=0$. The Kalb-Ramond field may be written as~\footnote{This 2-form differs from the one in~\cite{Minimal} 
and equation (\ref{Kalb}) below by an exact form whose pulled back counterpart neither contributes to the Euler-Lagrange equations nor to the on-shell regularized action.}
\be
B = - i \frac{q}{u^2}\dif v\wedge \dif \theta \ ,
\ee
where the flux mixing parameter is restricted to lie within $0\le q\le1$. The solution is~\cite{Minimal}
\be
\label{Solution}
\theta(\tau,\sigma)=\tau \ , \quad u(\tau,\sigma)=\sqrt{1-q^2}\sinh\sigma \ , \quad v(\tau,\sigma)=\textnormal{arctanh}\left(q\tanh\sigma\right) \ ,
\ee
where $\tau\in[0,2\pi)$ and $\sigma\in\left[0,\infty\right)$ are the Euclidean time and space world-sheet coordinates, respectively. We must emphasize that (\ref{Solution}) 
constitutes a particular limit of the general solution to the underlying integrable mechanical system, a feature that will justify the special properties that it presents as an effective background 
in the semiclassical quantization scheme. This solution induces a metric on the world-sheet $\Sigma$ for which $\tau$ and~$\sigma$ are isothermal coordinates,
\be
\label{Forbidden}
\dif  s^2_{\Sigma}=\frac{\dif \tau^2+\dif \sigma^2}{\left(1-q^2\right)\sinh^{2}\sigma} \ .
\ee
Its associated non-trivial component of the Riemann tensor and Ricci scalar are
\be
{R^{\tau}}_{\sigma\tau\sigma} = - \frac{1}{\sinh^2\sigma} \ , \quad R^{(2)} = - 2\left(1-q^2\right) \ .
\ee
Note that the coordinates $\tau$ and $\sigma$ are valid as long as $q \neq 1$, since otherwise the metric is singular. We can employ instead $\theta$ and $u$ as local world-sheet coordinates 
for which the metric is regular for arbitrary mixing parameter $q$. The singularity is still present, but it shows up in the relation between both coordinate systems in the limit of pure NS-NS flux. 
This feature is expected since the boundary of Euclidean $\AdS_{3}$ is conformal and thus it is located at an infinite distance from its center. Therefore, a proper parameterization 
of a world-sheet stuck therein is singular from the point of view of the bulk. The coordinates $\theta$ and $u$ display the metric non-conformally,
\be
\label{Induced}
\dif  s^2_{\Sigma} = \frac{1}{u^2}\left(\dif \theta^2 + \frac{\dif u^2}{1-q^2+u^2}\right) \ ,
\ee
and hence the non-trivial component of the Riemann tensor as
\be
{R^{\theta}}_{u\theta u}=-\frac{1-q^2}{u^2(1-q^2+u^2)} \ .
\ee
In the limit where the R-R flux vanishes the surface becomes locally flat, in accordance with the fact that it adheres to the boundary.

A surface with boundary conditions at the boundary of Euclidean $\AdS_{3}$ is linked with a divergent on-shell action. A regularization prescription for computing the latter is then needed. 
Here we will choose the scheme in which the boundary terms are removed. If we introduce an infrared regulator $\epsilon>0$ such that $\sigma\in\left(\epsilon,\infty\right)$, we find the on-shell action
\ba
\label{Action}
S & \!\! = \!\! & \sqrt{\lambda}\int_{\epsilon}^{\infty}\dif\sigma\frac{\cosh^2\sigma}{\sinh^2\sigma\left(\cosh^2\sigma-q^2\sinh^2\sigma\right)} \nonumber \\
& \!\! = \!\! & \sqrt{\lambda} \, \big( \cotanh\epsilon + q\arctanh q- q\arctanh\left(q\tanh\epsilon\right) -1 \big) \ ,
\ea
with $\lambda$ the 't Hooft coupling. Once the boundary terms are removed, the on-shell regularized action is
\be
S = \sqrt{\lambda} \, \big( q\arctanh q - 1 \big) \ .
\ee
This expression is valid as long as the R-R flux does not vanish. In the pure NS-NS flux limit, the world-sheet adheres to the boundary, and hence the associated regularized on-shell action vanishes after removing boundary terms. 

\subsection{Classical surface subtending two parallel lines}

We will now present the classical solution subtending two parallel lines. This surface may be expressed straightforwardly in the Cartesian parameterization of the Poincar\'e patch of Euclidean $\AdS_{3}$, with respect to which the metric is 
\be
\dif s^2 = \frac{\dif t^2+\dif x^2+\dif z^2}{z^2} \ ,
\ee
while the $B$-field is 
\be
\label{Kalb}
B = i \frac{q}{z^2}\dif t\wedge \dif x \ .
\ee
The solution reads
\be
t \left(\tau,\sigma\right) = \tau \ , \quad x\left(\tau,\sigma\right)=q\sigma \ , \quad z\left(\tau,\sigma\right)=\sqrt{1-q^2} \, \sigma \ ,
\ee
where $\tau\in\left(-\infty,\infty\right)$ and $\sigma\in\left[0,\infty\right)$. As in the previous case, the solution above induces a metric conformally 
on the world-sheet $\Sigma$,
\be
\label{cirteM}
\dif  s^2_{\Sigma}=\frac{\dif\tau^2+\dif\sigma^2}{\left(1-q^2\right)\sigma^2} \ ,
\ee
whose non-trivial components of the Riemann tensor and Ricci scalar are given by 
\be
{R^{\tau}}_{\sigma\tau\sigma}=-\frac{1}{\sigma^2} \ , \quad R^{(2)} = - 2\left(1-q^2\right) \ .
\ee
This is an Euclidean $\AdS_{2}$ surface embedded in Euclidean $\AdS_{3}$ confined to the boundary of the latter when the R-R flux vanishes. A regular coordinate system 
in the pure NS-NS flux limit may also be found anew. However, it will not be needed for our purposes. It is enough to note that the transformation between the latter and the isothermal coordinates 
is consistently singular in the limit of vanishing R-R flux. Finally, the on-shell action is not modified by the mixture of fluxes except for a global factor, and thus the regularization prescription 
and the consequent vanishing on-shell regularized action hold as in the $q=0$ limit. When $q=1$ the action is also zero since it becomes a boundary term.   

%%%%%%%%%%%%%%%%%%%%%%%%%%%%%%%%%%%%%%%%%%%%%%%%%%%%%%%%%%%%%%%%%%
%%%%%%%%%%%%%%%%%%%%%%%%%%%%%%%%%%%%%%%%%%%%%%%%%%%%%%%%%%%%%%%%%%

\section{Semiclassical partition function}

\label{soD}

In this section, we will derive the expression of the semiclassical partition functions associated to the solutions of the preceding section. 
We will employ the background field expansion, which provides us a quadratic Lagrangian in the perturbative fields that turns into a set of functional determinants of differential operators once the path integral 
is performed. We will then discuss the conformal anomaly and show that it entirely comprises the NS-NS flux contribution in the mixed flux regime. On the contrary, there is not such a term 
in the pure NS-NS flux limit, which requires a special treatment. We will postpone this case to section \ref{Neveu}, and focus the discussion below to the minimal surface subtending an annulus, 
arguing that the derivation for the strip case is almost identical. From now on we will refer to the surfaces subtending two concentric circumferences and two parallel lines at the boundary 
as the first and the second surfaces, respectively.

\subsection{Background field expansion}

The background field expansion consists in the study of quadratic perturbations around extrema of the classical action, which leads one to a Gaussian path integral in the perturbative fields. 
The perturbative bosonic Lagrangian may be obtained by expanding, around the classical solution, the Nambu-Goto action plus a Wess-Zumino term in the fluctuation fields  up to second order.
In this setting, it will be convenient to regard $u$ and $\theta$ as local world-sheet coordinates, and then fix the static gauge in which none of them is perturbed. 
The bosonic fluctuation fields should thus be taken in the $v$, the $\Esfera^{3}$ and the $\Toro^{4}$ directions. We will denote the first one by $\bar{v}$ 
and the remaining ones by $\bar{\xi}^{a}$, with $a=3,...,9$. These fields are supplied with vanishing Dirichlet boundary conditions at the boundary of Euclidean $\AdS_{3}$
and are further required to decay fast enough as they approach it. We will impose these conditions so that the integration by parts in the expansion is legitimated.

The second order Lagrangian for the bosonic fluctuations is then
\be
\label{Lagrangian}
{L}_{B}=\sqrt{h}h^{\alpha\beta}\partial_{\alpha}\bar{v}\partial_{\beta}\bar{v}+\sqrt{h}\left[R^{(2)}+4(1-q^2)\right]\bar{v}^2+\delta_{ab}\sqrt{h}h^{\alpha\beta}\partial_{\alpha}\bar{\xi}^{a}\partial_{\beta}\bar{\xi}^{b} \ ,
\ee
where we have rescaled the fluctuation field $\bar{v}$  as 
\be
\label{Redefinition}
\bar{v}\mapsto \frac{u}{\sqrt{1-q^2+u^2}} \, \bar{v} 
\ee
to write the Lagrangian in canonical form. We must stress that no component of the Kalb-Ramond field along $\Esfera^{3}$ enters the problem because it appears as an exact form, and thus it can 
be ignored by virtue of the aforementioned boundary conditions. The Lagrangian defines eight spectral problems, one per each fluctuation field. If $\Delta$ denotes the Laplace-Beltrami operator on $\Sigma$, 
the second order differential operators for $\bar{v}$ and $\bar{\xi}^{a}$ are, respectively,
\be
\label{Laplacian}
\D_{B,1}=-\Delta+R^{(2)}+4(1-q^2) \ , \quad \D_{B,2}=-\Delta \ .
\ee
Both spectral problems are supplemented with the norm derived from the inner product, 
\be
\label{Inner}
\bra{\varphi_{1}}\ket{\varphi_{2}}=\int_{\Sigma}\dif^{2}\sigma\sqrt{h}\ {{\varphi}_{1}}^{\! \! *} \varphi_{2} \ .
\ee
In fact, the inner product applies to $\bar{v}$ after being redefined by the factor $\sqrt{1+u^2}/u$ in the metric (\ref{Metric}). Note that such redefinition, 
together with the transformation by a factor $u/\sqrt{1-q^2+u^2}$ above, is expected to compensate the contribution of the ghost determinant arisen from the static 
gauge-fixing condition~\cite{Drukker}. This observation is indeed consistent with the explicit form of the latter, namely,
\be
\Delta_{\textnormal{ghost}}={{\det}^{1/2}\frac{1-q^2+u^2}{1+u^2}} \ .
\ee

Let us now move to the fermionic Lagrangian for the fluctuation fields. In order to proceed we should Wick-rotate back to $\AdS_{3}\times \Esfera^{3} \times \Toro^{4}$  
and introduce a zehnbein therein. After rotating back the angular coordinate through $\theta \mapsto i \theta$, the metric reads
\be
\dif s^2 = \eta_{ab}E^{a}E^{b} = - (E^{0})^2+(E^{1})^2+(E^{2})^2+\underset{a=3}{\overset{9}{\sum}}(E^{a})^{2} \ .
\ee
The dreibein of $\AdS_{3}$ is explicitly
\be
\label{Dreibein}
E^{0}=\frac{\dif\theta}{u} \ , \quad E^{1}=\frac{\dif u}{u\sqrt{1+u^2}} \ , \quad E^{2}=\frac{\sqrt{1+u^2}}{u}\, \dif v \ ,
\ee
and the spin connection hence is
\be
\Omega^{01}=-\Omega^{10}=-\sqrt{1+u^2} \, E^{0} \ , \quad \Omega^{12}=-\Omega^{21}=\frac{1}{\sqrt{1+u^2}}\, E^{2} \ . 
\ee
It will not be necessary to specify the siebenbein of $\Esfera^{3}\times\Toro^{4}$, or its associated spin connection, because once they are pulled back upon the 
world-sheet all of them vanish. On the contrary, both the R-R and NS-NS three-form fluxes are needed, 
\ba
H & \!\! = \!\! & -2q\left(E^{0}\wedge E^{1}\wedge E^{2}+E^{3}\wedge E^{4}\wedge E^{5}\right) \ , \nonumber  \\
F & \!\! = \!\! & -2\bar{q}\left(E^{0}\wedge E^{1}\wedge E^{2}+E^{3}\wedge E^{4}\wedge E^{5}\right) \ ,
\ea
where $\bar{q}=\sqrt{1-q^2}$. We will maintain the notation for differential forms after they are pulled back on the world-sheet since no confusion could arise.

The fermionic Lagrangian to quadratic order in the Green-Schwarz action is~\cite{Wulff} 
\be
\label{Density}
L_{F} = - i \big( \sqrt{-h}h^{\alpha\beta}\delta_{AB}-\epsilon^{\alpha\beta}(\sigma_{3})_{AB} \big) \bar{\Theta}^{A}\Gamma_{\alpha}{\left(\D_{\beta}\right)^{B}}_{C}\Theta^{C} \ ,
\ee
where $\Gamma_{\alpha}=E_{\alpha}^{a}\Gamma_{a}$. From (\ref{Dreibein}),
\be
\label{Uncanonical}
\Gamma_{\theta}=\frac{1}{u}\Gamma_{0} \ , \quad \Gamma_{u}=\frac{1}{u\sqrt{1+u^2}}\left(\Gamma_{1}-\frac{q}{\sqrt{1-q^2+u^2}}\Gamma_{2} \right) \ .
\ee
Besides, $\Theta^{A}$, with $A=1,2$, are sixteen-component Majorana-Weyl spinor, and the covariant derivatives are
\ba
{(\D_{\theta})^{A}}_{B} & \!\! = \!\! & \left(\partial_{\theta}-\frac{\sqrt{1+u^2}}{2u}\Gamma_{0}\Gamma_{1}\right){\delta^{A}}_{B} 
-\frac{1}{2u}\Gamma_{1}\Gamma_{2}\left(\bar{q}P\,{\left(\sigma_{1}\right)^{A}}_{B}+q\,{\left(\sigma_{3}\right)^{A}}_{B}\right) , \nonumber \\
{(\D_{u})^{A}}_{B} & \!\! = \!\! & \left(\partial_{u}+\frac{q}{2u(1+u^2)\sqrt{1-q^2+u^2}}\Gamma_{1}\Gamma_{2}\right) {\delta^{A}}_{B}\  \\
 & \!\! + \!\! & \frac{1}{2u\sqrt{1+u^2}}\Gamma_{0}\left(\frac{q}{\sqrt{1-q^2+u^2}}\Gamma_{1}-\Gamma_{2}\right)\left(\bar{q}P\,{\left(\sigma_{1}\right)^{A}}_{B}+q\,{\left(\sigma_{3}\right)^{A}}_{B}\right) \nonumber ,
\ea
where $P$ denotes the projection operator
\be
\label{Projector}
P = \frac{1}{2}\left(1+\Gamma_{0}\Gamma_{1}\Gamma_{2}\Gamma_{3}\Gamma_{4}\Gamma_{5}\right) \ .
\ee
In order to write this Lagrangian in a canonical two-dimensional covariant form we must perform a rotation. In a suitable kappa gauge, such a transformation allows one to derive a kinetic term 
akin to the one for two-dimensional world-sheet spinors. These steps permit the path integral to be carried out and thus the two-dimensional functional determinants to be obtained.
The rotation should transform $\Gamma_{\alpha}$ into the projection of the gamma matrices upon the world-sheet via the zweibein of the induced metric, namely,
\be
e^{0}=\frac{d\theta}{u} \ , \quad e^{1}=\frac{d u}{u\sqrt{1-q^2+u^2}} \ .
\ee
The rotation matrix $R=\exp\left(\varphi\Gamma_{1}\Gamma_{2}\right)$, with
\be
\cos\left(2\varphi\right)=\sqrt{\frac{1-q^2+u^2}{1+u^2}} \ , \quad \sin\left(2\varphi\right)=-\frac{q}{\sqrt{1+u^2}} \ ,
\ee
satisfies indeed the desired requirement, 
\be
\gamma_{\theta}=R^{-1}\Gamma_{\theta}R=\frac{1}{u}\Gamma_{0} \ , \quad \gamma_{u}=R^{-1}\Gamma_{u}R=\frac{1}{u\sqrt{1-q^2+u^2}}\Gamma_{1} \ .
\ee
If we also rotate the Majorana-Weyl spinors through $\Theta^{A}\mapsto R \, \Theta^{A}$  and further fix the kappa-symmetry with the condition $\Theta^{1}=\Theta^{2}\equiv\Theta$, we are led to the Lagrangian density
\be
\label{Simplified}
L_{F}=-2i\bar{\Theta}\sqrt{-h}\left(h^{\alpha\beta}\gamma_{\alpha}\nabla_{\beta}+\bar{q}\Gamma_{0}\Gamma_{1}\Gamma_{2}P\right)\Theta \ ,
\ee
where $\nabla_{\alpha}$ is the covariant derivative with respect to the induced metric, with components 
\be
\nabla_{\theta}=\partial_{\theta}-\frac{\sqrt{1-q^2+u^2}}{2u}\Gamma_{0}\Gamma_{1} \ , \quad \nabla_{u} =\partial_{u} \ . 
\ee
The operator in the Lagrangian (\ref{Simplified}) defines a spectral problem with an inner product analogous to that in (\ref{Inner}). 
Now we may diagonalize $P$ by performing an additional rotation. In this way, it is possible to split the Lagrangian into the sum of two densities, 
\ba
L_{F,1} & \!\! =  \!\! & 2\sqrt{-h}\,\bar{\theta}^{1}\D_{F,1}\theta^{1} = - 2i\sqrt{-h}\,\bar{\theta}^{1}\left(h^{\alpha\beta}\gamma_{\alpha}\nabla_{\beta} 
+ \bar{q}\Gamma_{0}\Gamma_{1}\Gamma_{2}\right)\theta^{1} \ , \nonumber  \\
L_{F,2} & \!\!=  \!\! & 2\sqrt{-h}\,\bar{\theta}^{2}\D_{F,2}\theta^{2} = - 2i\sqrt{-h}\,\bar{\theta}^{2}h^{\alpha\beta}\gamma_{\alpha}\nabla_{\beta}\theta^{2} \ , 
\label{Split}
\ea
where $\theta^{1}$ and $\theta^{2}$ are sixteen-component Majorana-Weyl spinors whose eigenvalues with respect to the projector $P$ are one and zero, respectively. 
If we finally Wick-rotate (\ref{Split}), we obtain the differential operators of interest. 

In sum, when the Gaussian path integration is performed, the subleading contribution to the semiclassical partition function is \footnote{We disregard the measure factors in the path integral that may be present.} 
\be
\label{Partition}
Z = \frac{\det^{2}\D_{F,1}\det^{2}\D_{F,2}}{\det^{1/2}\D_{B,1}\det^{7/2}\D_{B,2}} \ ,
\ee
and the one-loop effective action is
\be
\label{Effective}
\Gamma_{1} = - \log Z \ .
\ee

We noted at the beginning of this section that the steps in the derivation of the semiclassical partition function for the second surface are completely parallel. 
In this case we obtain an identical expression for the one-loop effective action, up to the substitution of the induced metric on the world-sheet of the first surface 
by the metric of the second one.

\subsection{Conformal anomaly and flat-measure determinants}

The spectral problem associated to the functional determinants under study is endowed with an inner product whose measure is defined through the induced metric upon the world-sheet. 
The latter is non-trivial whenever the R-R flux does not vanish as any of the two induced metrics is flat. In this case it is appropriate to factorize the semiclassical partition function into the product of a factor 
accounting for a conformal transformation, i.e., the conformal anomaly, and the ratio of determinants whose measure is flat. 

Let us consider the semiclassical partition function associated to the first surface. In order to proceed we need to find a transformation in which the induced metric is conformally flat. 
We may, for instance, choose the isothermal coordinates in (\ref{Solution}) (except in the regime of pure NS-NS flux). In those coordinates, the expression of the semiclassical partition function 
factorizes as $Z=Z_{q}\widehat{Z}$, where $Z_{q}$ is the conformal anomaly and $\widehat{Z}$ is the ratio of determinants with associated flat measure. It turns out that all the dependence on~$q$ is entirely ascribed to the conformal factor 
and that the expression of $\widehat{Z}$ is that of the pure R-R flux regime. A similar factorization holds when quadratic perturbations around the second surface are considered if the induced metric is written as (\ref{cirteM}). 

We will now focus on the conformal anomaly. Its finite remnant is given by the second Seeley coefficient of the heat kernel regularization scheme. 
This fact, together with other issues concerning the conformal anomaly, has been already discussed in~\cite{Drukker}. For a more thorough treatment of these topics we refer to~\cite{Drukker} and references therein. 
Here we just note that the fermionic functional determinants involve Green-Schwarz rather than world-sheet spinors, and hence the contribution of the former is four times larger than the one that the latter 
would provide~\cite{Kruczenski,Drukker}. If the count for the anomaly is performed, one finds that it is not zero, but a finite remnant is obtained. The upshot does not signal any inconsistency.
The Nambu-Goto and Polyakov path integrals are equivalent at one-loop~\cite{Fradkin}, but the cancellation of the conformal anomaly in the first case requires to extract a non-trivial contribution 
from its path integral measure. On the other hand, if spinors are dealt with as world-sheet spinors, the cancellation indeed occurs, in parallel with~\cite{Kruczenski, Cagnazzo}.

In order to find an explicit expression for $\widehat{Z}$, it is appropriate to render the fermionic determinants into the product of determinants of second order differential operators. 
This step appeals to the fact that $\Gamma_{0}\Gamma_{1}$ is anti-Hermitian, traceless and squares to minus the identity. In particular, we will introduce the factorization 
\be
{\det}^2 \D_{F,A} = \det \D_{F,A}^{+}\det\D_{F,A}^{-} \ , 
\ee
with $A=1,2$. Therefore, 
\be
\widehat{Z} = \frac{\det\widehat{\D}_{F,1}^{+}\det\widehat{\D}_{F,1}^{-}\det\widehat{\D}_{F,2}^{+}\det\widehat{\D}_{F,2}^{-}}{\det^{1/2}\widehat{\D}_{B,1}\det^{7/2}\widehat{\D}_{B,2}} \ .
\label{Trivial}
\ee
The differential operators involved in the expansion around the first surface are 
\ba
\widehat{\D}_{B,1} & \!\! = \!\! & - \left(\partial^{2}_{\tau}+\partial^{2}_{\sigma}\right)+2\cosech^2\sigma \ , \quad 
\widehat{\D}_{B,2} = - \left(\partial^{2}_{\tau}+\partial^{2}_{\sigma}\right) \ , \nonumber \\
\widehat{\D}_{F,1}^{\pm} & \!\! = \!\! & - \left(\partial^{2}_{\tau}+\partial^{2}_{\sigma}\right)\pm i\cotanh\sigma\partial_{\tau}+\frac{3}{4}\cosech^2\sigma+\frac{1}{4} \ , \label{Operators} \\ 
\widehat{\D}_{F,2}^{\pm} & \!\! = \!\! & - \left(\partial^{2}_{\tau}+\partial^{2}_{\sigma}\right)\pm i\cotanh\sigma\partial_{\tau}-\frac{1}{4}\cosech^2\sigma+\frac{1}{4} \ , \nonumber
\ea
while those involved in the expansion around the second surface are
\ba
\widehat{\D}_{B,1} & \!\! = \!\! & - \left(\partial^{2}_{\tau}+\partial^{2}_{\sigma}\right)+\frac{2}{\sigma^2} \ , \quad 
\widehat{\D}_{B,2} = - \left(\partial^{2}_{\tau}+\partial^{2}_{\sigma}\right) \ , \nonumber \\
\widehat{\D}_{F,1}^{\pm} & \!\! = \!\! & - \left(\partial^{2}_{\tau}+\partial^{2}_{\sigma}\right)\pm \frac{i}{\sigma}\partial_{\tau}+\frac{3}{4\sigma^2} \ \label{srotarepO} , \\ 
\widehat{\D}_{F,2}^{\pm} & \!\! = \!\! & - \left(\partial^{2}_{\tau}+\partial^{2}_{\sigma}\right)\pm \frac{i}{\sigma}\partial_{\tau}-\frac{1}{4\sigma^2} \ . \nonumber
\ea

%%%%%%%%%%%%%%%%%%%%%%%%%%%%%%%%%%%%%%%%%%%%%%%%%%%%%%%%%%%%%%%%%%
%%%%%%%%%%%%%%%%%%%%%%%%%%%%%%%%%%%%%%%%%%%%%%%%%%%%%%%%%%%%%%%%%%

\section{Functional determinants}

\label{serT}

We will compute now the functional determinants of the second-order differential operators put forward in (\ref{Operators}) and (\ref{srotarepO}) by means of the Gel'fand-Yaglom method~\cite{Dunne}. 
This technique is applicable if the operators under consideration are one-dimensional. In our case, we can reduce the operators in (\ref{Operators}) and (\ref{srotarepO}) to ordinary differential operators 
invoking the boundary conditions for the fields with respect to the Euclidean time world-sheet coordinate. This option is available as a consequence of the usage of the ansatz that renders the classical setting 
into an integrable system, which leads to a $\tau$-independent effective two-dimensional background in the semiclassical quantization scheme.

We will start the analysis with the first surface. The fluctuation fields are periodic in time for the bosons, and anti-periodic for the fermions~\cite{Kruczenski}. 
Therefore, a decomposition in Fourier modes is allowed for all of them. One may thus write $\widehat{Z}$ in terms of $\widehat{Z}_{n}$, where $\widehat{Z}_{n}$ is the $n$-th mode ratio of determinants. 
The latter incorporates the ratio of one-dimensional determinants whose operators are obtained from equations (\ref{Operators}) through the replacements $\partial_{\tau}\mapsto-in$ for $\widehat{\D}_{B,1}$ 
and $\widehat{\D}_{B,2}$, $\partial_{\tau}\mapsto-i\left(n-1/2\right)$ for $\widehat{\D}_{F,1}^{+}$ and $\widehat{\D}_{F,2}^{+}$, and $\partial_{\tau}\mapsto-i\left(n+1/2\right)$ for $\widehat{\D}_{F,1}^{-}$ and $\widehat{\D}_{F,2}^{-}$. 
After introducing these replacements the determinants read
\ba
\label{molgaY}
\widehat{\D}_{B,1} & \!\! = \!\! & - \partial^{2}_{\sigma}+n^{2}+2\cosech^{2}\sigma \ , \quad
\widehat{\D}_{B,2} = - \partial^{2}_{\sigma}+n^{2} \ , \nonumber \\
\widehat{\D}_{F,1}^{\pm} & \!\! = \!\! & - \partial^{2}_{\sigma}+\left(n\mp\frac{1}{2}\right)^2\pm\left(n\mp\frac{1}{2}\right)\cotanh\sigma+\frac{3}{4}\cosech^2\sigma+\frac{1}{4} \ , \\ 
\widehat{\D}_{F,2}^{\pm} & \!\! = \!\! & - \partial^{2}_{\sigma}+\left(n\mp\frac{1}{2}\right)^2\pm\left(n\mp\frac{1}{2}\right)\cotanh\sigma-\frac{1}{4}\cosech^2\sigma+\frac{1}{4} \ . \nonumber
\ea
Regarding the fermionic operators, we must stress that the conversion of the sum over half-integers into a sum over integers involves an intermediate regularization process 
whose remnant is decisive for the cancellation of the infrared divergences. We will study this issue below once we resort to the Gel'fand-Yaglom method, which we will employ to obtain 
an expression for $\widehat{Z}_{n}$, and the Abel-Plana formula, on the basis of which we will address the difference of the one-loop effective actions. 
Moreover, the symmetry property $\widehat{Z}_{n} = \widehat{Z}_{-n}$ enables us to express 
\be
\label{amuS}
\log\widehat{Z}=\log\widehat{Z}_{0}+2\,\overset{\infty}{\underset{n=1}{\sum}}\,\log\widehat{Z}_{n} \ .
\ee

Let us consider now the second surface. In this case, fields are neither periodic nor anti-periodic concerning their time dependence.~\footnote{Equivalently, one may take the Euclidean time world-sheet to be periodic and make the period tend to infinity.} 
However, as the time interval now is non-compact, we may perform a continuous Fourier transform, according to which $\widehat{Z}$ can be expressed in terms of $\widehat{Z}_{p}$, 
where each $\widehat{Z}_{p}$ consists of the ratio of determinants derived from (\ref{srotarepO}) by means of the substitution of $\partial_{\tau}\mapsto-ip$. The operators now read
\be
\label{Reference}
\begin{split}
\widehat{\D}_{B,1}& = - \partial^{2}_{\sigma}+p^{2}+\frac{2}{\sigma^2} \ , \quad
\widehat{\D}_{B,2} = - \partial^{2}_{\sigma}+p^{2} \ , \\
\widehat{\D}_{F,1}^{\pm} & = - \partial^{2}_{\sigma}+p^{2}\pm \frac{p}{\sigma}+\frac{3}{4\sigma^2} \ , \\ 
\widehat{\D}_{F,1}^{\pm} & = - \partial^{2}_{\sigma}+p^{2}\pm \frac{p}{\sigma}-\frac{1}{4\sigma^2} \ , \\ 
\end{split} 
\ee
and the symmetry property $\widehat{Z}_{p}=\widehat{Z}_{-p}$ leads to
\be
\label{Manipulation}
\log \widehat{Z} = 2 \int_{0}^{ \infty}\log\widehat{Z}_{p} \ .
\ee

As we have already stated, in the appendix~\ref{Yaglom} we will apply the Gel'fand-Yaglom method to obtain the expressions for the determinants. 
The method reduces the computation to the attainment of a solution to an initial value problem. However, in order to apply the method it is necessary to shift the boundary values 
of the problem from $\sigma=0$ and $\sigma=\infty$ to $\sigma=\epsilon$ and $\sigma=R$, respectively. The former point acts as an infrared regulator, bringing the boundary of the classical solution 
to a finite distance from the center of Euclidean AdS$_{3}$. Moreover, it is needed since otherwise some potential terms of (\ref{molgaY}) and (\ref{Reference}) would be singular at the leftmost boundary. 
The latter point is introduced as a computational device, in such a way that the (possibly divergent) outcome corresponds to the $R\rightarrow\infty$ limit. In fact, the criterion on the basis 
of which we associate one and the same $n$ (or $p$ for the second surface) to these different operators refers to the asymptotic behaviour in the large $R$ regime of their individual determinants, 
so that the dependence on the regulator of the resultant ratio $\widehat{Z}_{n}$ (respectively $\widehat{Z}_{p}$) proves to be erased. \footnote{The coefficients of the subleading terms in the asymptotic expansion are of the same, 
or less, order in either $n$ or $p$ than those of the leading terms, and hence they are negligible in the large $R$ limit.}

The resolution of the initial value problem underlying the computation of the determinant of each individual differential operator 
is relegated to the appendix \ref{Yaglom}.~\footnote{In reference~\cite{David} other boundary conditions for massless fermionic spectral problems than the ones imposed here have been introduced to apply the Gel'fand-Yaglom method. 
Their choice is motivated by the comparison with the result obtained in the dual side of the correspondence that we lack here.} Since the dependence on the upper bound regulator cancels, as we have stated above, the $R\rightarrow\infty$ limit can be taken. 
In the first case the expression of the $n$-th factor of the semiclassical partition function in (\ref{amuS}) with $n>0$ is~\footnote{In fact, to arrive to this expression 
we have made a shift of the mode number $\log n\mapsto\log \left(n+1\right)$ in one determinant. The sum over Fourier modes is of course insensitive to this operation.}
\be
\label{Number}
\widehat{Z}_{n} = \sqrt{\frac{n+1}{n+\cotanh\epsilon}} \, \big[ \left(2n+1\right)\sinh\epsilon+\cosh\epsilon \big] \e^{\left(2n-1\right)\epsilon}\B (\e^{-2\epsilon};n,0) \ ,
\ee
with $\B(x;a,b)$ the incomplete Euler beta function, whereas for $n=0$ it is
\be
\label{Cero}
\widehat{Z}_{0}= \sqrt{\tanh\epsilon} \ .
\ee
On the other hand, the $p$-th term in the integrand of (\ref{Manipulation}) is 
\be
\label{Recomposition}
\widehat{Z}_{p} = \sqrt{\frac{p\epsilon}{p\epsilon+1}}\left(2 p \epsilon+1\right)\e^{2p\epsilon}\Ei_{1}(2p\epsilon) \ ,
\ee
where $\Ei_{1}(x)$ is the exponential integral. 

\subsection{Finiteness of the ratio}

We will now follow~\cite{Kristjansen} and show that the ratio between the semiclassical partition functions is finite by representing the sum (\ref{amuS}) as a divergent integral plus a remnant through the Abel-Plana formula, 
\be
\label{Formula}
\overset{\infty}{\underset{n=1}{\sum}}\,f(n) = - \frac{1}{2}f(0) + \int_{0}^{\infty} \dif x f(x)+i\int_{0}^{\infty}\dif x \, \frac{f(ix)-f(-ix)}{\e^{2\pi x}-1} \ .
\ee
The cases to which it is applied are arrayed in appendix~\ref{Plana}. We must emphasize that in our context the formula should be viewed as formal, since $\widehat{Z}_{n}$ 
does not satisfy the requirements that legitimate its employment. Nonetheless, the result is finite once the contribution of the second surface is subtracted and the limit $\epsilon \rightarrow 0$ is taken in the integrand.

It is proved in appendix \ref{Plana} that the latter term in the Abel-Plana formula is always real and infrared finite for the cases at hand (see equations (\ref{Lineal}), (\ref{Gradshteyn}) and (\ref{Beta})). 
Therefore, the divergence is ascribed to the first two terms.  We will consider first the divergent integral term. According to the previous discussion, we have to subtract from it the term in (\ref{Recomposition}). 
Firstly, we split here the difference between integral expressions into
\be
\label{Integral}
I = 2\int_{0}^{\infty}\dif x\left[\log \left(\sqrt{\frac{ x\epsilon +1}{\left(x+\cotanh\epsilon\right)\epsilon}}\frac{\left(2x+1\right)\sinh\epsilon+\cosh\epsilon}{2 x\epsilon +1}\frac{\B\left(\e^{-2\epsilon};x+1,0\right)}
{\Ei_{1}(2 x\epsilon)}\right)+\epsilon\right] + 1 \ .
\ee
The integral has been derived rewriting the integral of the logarithm of the term in the numerator of the square root in (\ref{Recomposition}) like
\be
\label{Escindido}
\int_{0}^{\infty}\dif x\log \sqrt{x}=\int_{0}^{\infty}\dif x\log\sqrt{x+1}-1/2 \ ,
\ee
which is defined when considered into the ratio of semiclassical partition functions. 

Although the operations of taking the limit of vanishing infrared regulator and performing the integral do not commute, it is more plausible that the former precedes the latter. 
This can be justified by the fact that $\epsilon$ has been introduced as a regulator, which allows one to deal with each of the two sets of functional determinants independently. 
However, the well-defined object is their ratio, due to the expected cancellation of infrared divergences. Therefore, we can disregard the regulator 
when the two sets are paired and thus to deal with the ratio when $\epsilon\rightarrow0$ directly.~\footnote{An argument supporting the order of the limits chosen here is provided in~\cite{Pando} in a related context.} 
If we apply the limit to both integrands and employ the asymptotic expansions
\ba
\B(z;x,0) & \!\! = \!\! & -\log\left(1-z\right)-\gamma-\psi\left(z\right) + \left(1-x\right)\left(z-1\right)+\mathcal{O}\left(\left(z-1\right)^2\right) \ , \nonumber \\
\Ei_{1}(z) & \!\! = \!\! & -\log z-\gamma+z+ \mathcal{O}\left(z^2\right) \ ,
\ea
we conclude that $I \rightarrow 1$. Besides, the contribution of the first term of the Abel-Plana formula in (\ref{Formula}) plus $\widehat{Z}_{0}$ is
\be
\label{One}
I_{0}=-\epsilon+\log\B (\e^{-2\epsilon};1,0) \ ,
\ee
which diverges in the $\epsilon\rightarrow0$ limit. It diverges because it has to be paired with the remnant of the regularization procedure relating the sums over half-integers and integers. 

\subsection{Regularization of the sum over half-integer modes}

We have noted below equation (\ref{molgaY}) that the conversion of the sum over half-integers into the sum over integers of fermionic modes involves an intermediate 
regularization process which provides an infrared divergent, but otherwise finite, remnant. Such a regularization has been used by the authors of~\cite{Kruczenski} 
following reference~\cite{Frolov}. In particular, prior to arriving to the sum over integers, one should address the sum
\be
\label{Initial}
S = \underset{n\in\mathbb{Z}+1/2}{\sum}\Omega_{n}^{1} \ ,
\ee
which accounts for the fermionic contribution to the one-loop effective action. The regularization procedure consists in redefining the sum via an exponential suppression, 
~\footnote{The bosonic contribution is regularized analogously. We omit it here for the sake of conciseness.}
\ba
\label{Supersymmetric}
S_{1} & \!\! = \!\! & \lim_{\mu\rightarrow 0}\underset{n\in\mathbb{Z}+1/2}{\sum}\e^{-\mu\left|n\right|}\Omega_{n}^{1}
= \frac{1}{2}\lim_{\mu\rightarrow 0}\overset{\infty}{\underset{n=-\infty}{\sum}}\e^{-\mu\left|n\right|}\left(\Omega_{n-1/2}^{1}+\Omega_{-n-1/2}^{1}\right) \nonumber \\
& \!\! + \!\! & \lim_{\mu\rightarrow 0}\overset{\infty}{\underset{n=1}{\sum}}\e^{-\mu n}\left[\big(\e^{\frac{1}{2}\mu}-1\big)\Omega_{n-1/2}^{1}+\big(\e^{-\frac{1}{2}\mu}-1\big)\Omega_{-n-1/2}^{1}\right] \ .
\ea 
It is worth to point out that there is sharp difference in the motivation behind the introduction of the prescription in references~\cite{Kruczenski} and~\cite{Frolov}. In reference~\cite{Frolov} 
it is employed in such a way that the first line in the last equality above is finite when its regularized bosonic counterpart is added, and the second provides a finite remnant. 
In reference~\cite{Kruczenski}, the first line, plus the bosonic contribution, diverges in the same way as the initial sum (\ref{Initial}), whereas the second provides the aforementioned infrared divergent term. 
Such a term is not unessential, since it crucially enters in the cancellation of infrared divergences. Taking this observation into account, it is then sensible to look upon the regularization prescription 
for the scenario we are considering as a procedure which allows us to derive correct infrared divergent terms, in addition to possible infrared finite remnants. 
Accordingly, one may introduce a shift on the regularization prescription exponentials above as long as it leads to a finite result. 
This is the case for the sum over half-integer modes in connection with the last operators listed in (\ref{molgaY}), for which we can regularize the sum as
\ba
\label{Shift}
S_{2} & \!\! = \!\! & \lim_{\mu\rightarrow 0}\underset{n\in\mathbb{Z}+1/2}{\sum}\e^{-\mu\left|n+1\right|}\Omega_{n}^{1}=\frac{1}{2}\lim_{\mu\rightarrow 0}\overset{\infty}{\underset{n=-\infty}{\sum}}\e^{-\mu\left|n\right|}
\left(\Omega_{n-1/2}^{1}+\Omega_{-n-1/2}^{1}\right) \nonumber \\
& \!\! + \!\! & \lim_{\mu\rightarrow 0}\overset{\infty}{\underset{n=1}{\sum}}\e^{-\mu n}\left[\big(\e^{\frac{1}{2}\mu}-1\big)\Omega_{-n-1/2}^{1}+\big(\e^{-\frac{1}{2}\mu}-1\big)\Omega_{n-1/2}^{1}\right] \ .
\ea
We must stress that whereas the first contribution is the same as the one for $S_{1}$, the second differs with respect to its counterpart. 
The expression of the remnant in both sums may be computed by successive application of the Gel'fand-Yaglom method and the Abel-Plana formula.

We will now compute the remnant for the first regularization. According to equations (\ref{Uno}) and (\ref{Dos}), 
we have to consider, at the first significant order in $\mu$, 
\be
S_{1} = \lim_{\mu\rightarrow 0}\overset{\infty}{\underset{n=1}{\sum}}\,\mu\e^{-\mu n}\left[\log\frac{n+1}{(2n+1)\sinh\epsilon+\cosh\epsilon}-\epsilon\right] \ ,
\ee
which, by means of the Abel-Plana formula, becomes,
\be
\label{Above}
\begin{split}
S_{1} & = \lim_{\mu\rightarrow 0}\int_{0}^{\infty}\dif x\,\mu\e^{-\mu x}\left[\log\frac{x+1}{(2x+1)\sinh\epsilon+\cosh\epsilon}-\epsilon\right]=-\log\left(2\sinh\epsilon\right)-\epsilon \ .
\end{split}
\ee
We note here that we can disregard the first and the last parts of (\ref{Formula}) since they are finite in $\mu$. 
Furthermore, since this limit yields non-vanishing zeroth order terms in $\mu$, higher order terms in the expansion that the parentheses of (\ref{Supersymmetric}) comprise vanish.

We will compute now the remnant of the second regularization (\ref{Shift}). Because of the absence of divergent terms in when $\mu\rightarrow0$ in $S_{1}$, 
the limit cannot diverge for the prescription to be congruent. The equations that we should take into account here are (\ref{Tres}) and (\ref{Cuatro}), 
and hence the sum, to the first significant order in $\mu$, is
\be
\label{Sum}
S_{2} = - \lim_{\mu\rightarrow 0}\overset{\infty}{\underset{n=1}{\sum}}\,\mu\e^{-\mu n}\left[\log\big( n \B (\e^{-2\epsilon};n,0) \big) + 2 n\epsilon\right] \ ,
\ee
that can be written through the Abel-Plana formula as
\ba
\label{Segunda}
S_{2} & \!\! = \!\! & -\lim_{\mu\rightarrow 0}\int_{1}^{\infty}\dif x\,\mu\e^{-\mu x}\left[\log\big(x\B (\e^{-2\epsilon};x,0) \big)+2x\epsilon \right] \nonumber \\
& \!\! = \!\! & -\lim_{\mu\rightarrow 0}\left[\frac{2\epsilon}{\mu}+\Ei_{1}\left(\mu\right)+\int_{1}^{\infty}\dif x\mu\e^{-\mu x}\log\B\left(\e^{-2\epsilon};x,0\right)\right] \ .
\ea
We have not succeeded in deriving an explicit analytic expansion for the last term above. Nevertheless, we can show that its divergence in $\mu$ is cancelled by that of its two previous ones. 
The reasoning relies on the argument according to which the non-negligible part of the integrand comes from the high region of integration, where its contribution is comparable with the $\mu\rightarrow0$ limit. 
It is thus accurate enough to resort to the asymptotic expansion~\cite{Sesma}
\be
\label{Expansion}
\B(\e^{-2\epsilon};x,0)\sim\frac{\e^{-2 \epsilon x}}{\left(1-\e^{-2\epsilon}\right)x}\overset{\infty}{\underset{n=0}{\sum}}\frac{a_{n}}{x^{n}} \ , \quad x\rightarrow+\infty \ ,
\ee
where $a_{n}$ are some coefficients independent of $x$ with $a_{0}=1$. Therefore,
\be
\label{Logarithm}
\log\B(\e^{-2\epsilon};x,0) = -2\epsilon x-\log x+\mathcal{O}\left(1\right) \ .
\ee
Now, if we take into account the relations (with $\mu>0$),
\be
\label{Terms}
\int_{1}^{\infty}\dif x\,x\e^{-\mu x}=\frac{1}{\mu}\left(1+\frac{1}{\mu}\right)\e^{-\mu} \ , \quad \int_{1}^{\infty}\dif x \log x \e^{-\mu x}=\frac{\Ei_{1}\left(\mu\right)}{\mu} \ , 
\ee
and that the integration of the terms omitted in the expansion of the logarithm are finite when $\mu\rightarrow0$, we find that the first two terms in (\ref{Segunda}) are indeed cancelled with the ones above. 
But we still need to find the remnant that allows the infrared terms to be cancelled in the partition function. We may argue that the terms of zeroth order are those appropriate to cancel the one in (\ref{One}) 
by studying the contribution coming from the upper and lower endpoints of integration of (\ref{Segunda}). Let us first consider the upper endpoint of integration. If we employ the full asymptotic expansion, we find that
\be
\label{Asymptotic}
S_{2} \sim \log\left(1-\e^{-2\epsilon}\right) \ ,
\ee
since the contribution of terms of order $\mathcal{O}(1/x)$ in the asymptotic series vanishes. This term cancels the infrared divergence emerging in $S_{1}$ 
(here $\sim$ refers that the asymptotic expansion has been employed to compute the sum). Nevertheless, the expansion misses any contribution from the small $x$ region, 
which could not be ignored, as shown by the change of variables $x\mapsto x/\mu$.
Therefore, if we integrate by parts, we obtain
\be
\label{Parts}
\int_{1}^{\infty}\dif x \mu \e^{-\mu x} \log\B (\e^{-2\epsilon};x,0) = \e^{-\mu}\log\B (\e^{-2\epsilon};1,0) +\int_{1}^{\infty}\dif x \e^{-\mu x} \partial_{x}\log \B (\e^{-2\epsilon};x,0) \ .
\ee
We note that the first term is not expected to be attainable if the asymptotic series is applied beforehand, since it comes from the lower integration region, as we have already discussed. 
In fact, this term cancels the infrared divergent contribution of (\ref{One}). The latter term is expected to account for (\ref{Asymptotic}) up to some infrared finite terms. 
Note that the first part of the expression above cannot reproduce it due to its different asymptotic behavior in the limit of vanishing infrared regulator.

In sum, if we add all the potentially divergent terms, we obtain an infrared finite result. The final result, however, is still ambiguous by potentially infrared finite terms emerging from $S_{2}$. 
If we assume that there are no such terms, the limit for the difference of one-loop effective actions, denoted again by $\Gamma_{1}$, would be
\be
\label{Resultado}
\Gamma_{1}=-\frac{1}{2}\log\left(2\pi\right) \ ,
\ee
where we have added all terms obtained in this section and in appendix \ref{Plana} when $\epsilon\rightarrow0$.

We may use the Riemann zeta-function regularization approach to support the previous statement. The method has been reviewed exhaustively in the literature, and hence we will not present it here (see for instance~\cite{Fursaev} 
for a reference on the subject). In this scheme, one does not need to resort to any reference solution and thus the semiclassical partition function of the solution subtending a circle at the boundary may be addressed directly. 
We may then borrow the formulae presented in~\cite{Aguilera} and apply them directly to the case considered here. If we were to proceed in this way, we would obtain again the result in~(\ref{Resultado}). 
Of course, the equivalence requires that the partition function of the reference solution trivializes up to one-loop order as in the Euclidean $\AdS_{5}\times \Esfera^{5}$ 
scenario.~\footnote{Even if both procedures seem to agree in our problem, this is not the case in general \cite{Giangreco,Pando,Kim,Aguilera}.}

%%%%%%%%%%%%%%%%%%%%%%%%%%%%%%%%%%%%%%%%%%%%%%%%%%%%%%%%%%%%%%%%%%
%%%%%%%%%%%%%%%%%%%%%%%%%%%%%%%%%%%%%%%%%%%%%%%%%%%%%%%%%%%%%%%%%%

\section{The limit of pure NS-NS flux}

\label{Neveu}

In section~\ref{soD} we noted that the pure NS-NS flux limit requires a separate treatment. Indeed, the isothermal coordinates employed in the previous sections 
are singular in the limit of pure NS-NS flux (see equations (\ref{Forbidden}) and (\ref{cirteM})). Therefore, we should employ a regular coordinate system 
that can be obtained from the isothermal coordinates through a singular change of variables. For instance, we may use the one that brings the metric 
in~(\ref{Forbidden}) into the form~(\ref{Induced}) for the first surface, and an analogous transformation for the second one. The change of variables shows that the metric becomes flat 
when the R-R flux vanishes, because the minimal surface is confined to the boundary of the Euclidean anti-de Sitter space. Furthermore, Dirichlet boundary conditions 
cannot be imposed to the perturbative fields at this boundary because if we intend to maintain Dirichlet conditions the problem is not well-defined. 
This is a consequence of the absence of non-vanishing fields over which the path integral could be performed. In order to solve this problem, we will assume 
that we can define a semiclassical partition function in this setting. We are thus led to two flat two-dimensional problems, one with eight free bosonic functional determinants and the other with eight free fermionic ones. 
Then, we impose asymptotic Dirichlet boundary conditions at the endpoints of the spatial interval.~\footnote{Dirichlet boundary conditions are again admissible in this context according to general arguments~\cite{Fradkin}.} 
In fact, we may consider a finite interval for the annulus surface, thereby extending the computation to every solution that belongs to the general class of world-sheets with vanishing dilatation charge 
in the limit of pure NS-NS flux~\cite{Minimal} (for simplicity we set the winding number to $k=1$ in this discussion). If we now successively apply the Gel'fand-Yaglom method and the Abel-Plana formula, 
we find that no infrared regulator is needed. In the case of the first surface we conclude that the one-loop effective action equals the squared length of the interval, 
due to the anti-periodic boundary conditions for fermions, whereas for the second surface it vanishes.  

This distinctive feature does not restrictively concern the semiclassical partition function of the solutions considered here. Indeed, the extension under the presence of NS-NS 
flux of the minimal surface termed ``quark-antiquark potential'' in the context of the AdS$_{5}/$CFT$_{4}$ correspondence, which subtends two parallel lines at the boundary of Euclidean AdS$_{3}$, 
can be shown to adhere to the boundary in one of the two limits in which the R-R flux vanishes (that is, $q=1$ or $q=-1$, depending on the conventions).~\footnote{The extension refers to 
the usage of the usual ansatz in the Wess-Zumino term accounting for the $B$-field. The pure NS-NS flux limit of this solution has been briefly studied in~\cite{Sonnenschein}.} 
Similarly, the class of minimal surfaces subtending two concentric circumferences at the boundary of Euclidean AdS$_{3}$ considered in~\cite{Minimal} displays a range of parameters 
for which the confinement of the world-sheet to the boundary in the limit of pure NS-NS flux again occurs. We may proceed analogously in these generalized cases, 
although the study of quadratic perturbation around those solutions and their associated functional determinants is considerably more involved. In particular, the contribution of the NS-NS flux in the functional determinants 
is not longer factorizable in the conformal anomaly. However, each problem reduces to one of the previously considered cases in the pure NS-NS flux limit, where minimal surfaces are stuck at the boundary 
of Euclidean AdS$_{3}$, and hence all the analysis and the conclusions above also apply for them. 

%%%%%%%%%%%%%%%%%%%%%%%%%%%%%%%%%%%%%%%%%%%%%%%%%%%%%%%%%%%%%%%%%%
%%%%%%%%%%%%%%%%%%%%%%%%%%%%%%%%%%%%%%%%%%%%%%%%%%%%%%%%%%%%%%%%%%

\section{Conclusions}

\label{Conclusions}

In this article we have studied the difference between one-loop effective actions of the extension under mixed R-R and NS-NS three-form fluxes of the minimal surfaces subtending 
a circle and a line at the boundary of Euclidean anti-de Sitter space. We have first presented their classical world-sheets and we have shown that they are confined to the boundary in the limit of pure NS-NS flux. 
We have then employed the background field expansion to obtain the functional determinants corresponding to each surface. We have found that the regimes of mixed flux and of pure NS-NS flux should be regarded separately. 
In the first case, we have encountered that the NS-NS flux contribution is comprised in the conformal anomaly for both surfaces. Therefore, the flat-measure determinants stand the same as in the pure R-R flux limit. 
In order to compute the contribution in the regime of pure R-R flux we have exploited the boundary conditions of the Euclidean time to express each flat-measure determinant 
as an infinite product of one-dimensional determinants. We have applied successively the Gel'fand-Yaglom method and the Abel-Plana formula to argue that the ratio of semiclassical partition functions 
associated to each surface is infrared finite. In doing so, we have employed that the conversion of the sum of half-integers to integers requires a shifted regularization prescription for massless fermionic determinants. 
Finally, we have analyzed the pure NS-NS flux limit of the setting. We have discussed the trivialization of the one-loop effective action in this regime. 
We have also shown that the reasoning applies to other surfaces that become stuck at the boundary of Euclidean AdS$_{3}$.  

The most immediate extension of our analysis concerns the computation of the one-loop effective action, along the lines of~\cite{Dekel}, of the whole family of minimal surfaces subtending two concentric circumferences 
at the boundary deformed under mixed fluxes. They present a region of parameters for which the world-sheet subtends an annulus whose outer radius diverges in the pure NS-NS flux limit. 
The role of reference solution, by means of which the infrared divergences are removed, could be played by the mixed flux deformation of the ``quark-antiquark potential'' surface, 
that adheres to the boundary in one of the two possible limits of pure NS-NS flux, as we have discussed in section~\ref{Neveu}. However, in the complementary limit it is located in the bulk 
and it subtends two infinitely distant lines at the boundary. It would be instructive to obtain the pure NS-NS flux limit of the ratio of semiclassical partition functions in this context and, specifically, 
find out which simplification occurs, if any, for those surfaces which do not get stuck at the boundary. 

A correlative problem is the extension of the classical world-sheet to a non-trivial latitude in the sphere and the computation of the difference of its one-loop effective action with the one for the solution considered here. 
Such a quantity has been derived in the realizations of type IIB string theory in $\AdS_{5}\times \Esfera^{5}$ and $\AdS_{4}\times\mathbb{CP}^{3}$ backgrounds starting from various approaches. 
In particular, the application of the phase shift method~\cite{Factor,Cagnazzo} has allowed to overcome the previously existing discrepancies, and match finally the field theory prediction. 
Although the dual picture is not manifest in the $\AdS_{3}\times \Esfera^{3}\times\Toro^{4}$ background as opposed to those higher dimensional scenarios, the application of the procedure should be still legitimate, 
and hence its upshot is expected to be again trustable. \footnote{This problem would require to employ a diffeomorphism-invariant regulator instead of the upper bound $R$ considered here.} 

Another path that may be worth pursuing concerns the relationship between the results presented here and D1-strings. In reference~\cite{Minimal}, 
the class of F1-strings were paired with that of D1-strings by means of the S-duality symmetry of type IIB string theory. In fact, D1-strings were shown to be the only configurations 
that could be connected through this symmetry with the minimal surfaces at hand. Such a connection entails, in particular, that the behavior of F1-strings in the pure NS-NS flux regime 
is mimicked by D1-strings in the limit of pure R-R flux. Since S-duality is non-perturbative in character, the connection between these configurations should persist beyond the leading order 
in the strong coupling regime. It would be thus interesting to find how the peculiarities arising in the limit of pure NS-NS flux of our setting at one-loop are reproduced in the D1-string picture.

%%%%%%%%%%%%%%%%%%%%%%%%%%%%%%%%%%%%%%%%%%%%%%%%%%%%%%%%%%%%%%%%%%
%%%%%%%%%%%%%%%%%%%%%%%%%%%%%%%%%%%%%%%%%%%%%%%%%%%%%%%%%%%%%%%%%%

\vspace{12mm}

\centerline{\bf Acknowledgments}

\vspace{2mm}

\no
The work of R. H. is supported by grant PGC2018-095382-B-I00 and by BSCH-UCM through grant GR3/14-A 910770. 
The work of J.~M.~Nieto is supported by the EPSRC-SFI grant EP/S020888/1 \emph{Solving Spins and Strings}. 
R.~Ruiz acknowledges the support of the Universidad Complutense de Madrid through the predoctoral grant CT42/18-CT43/18.

\appendix

\renewcommand{\theequation}{\thesection.\arabic{equation}}
\csname @addtoreset\endcsname{equation}{section}

%%%%%%%%%%%%%%%%%%%%%%%%%%%%%%%%%%%%%%%%%%%%%%%%%%%%%%%%%%%%%%%%%%
%%%%%%%%%%%%%%%%%%%%%%%%%%%%%%%%%%%%%%%%%%%%%%%%%%%%%%%%%%%%%%%%%%

\section{The Gel'fand-Yaglom method}

\label{Yaglom}

The Gel'fand-Yaglom method provides a way to compute functional determinants eluding any explicit reference to the eigenvalues of the ordinary differential operator of interest. 
Specifically, we will consider the regular second order differential operator $\mathcal{O}$ defining a Sturm-Liouville problem over a finite interval $[a,b]$ with Dirichlet boundary conditions on the endpoints, 
which is the case that encompasses all the spectral problems of interest for the purposes of the article. The Gel'fand-Yaglom method states that the determinant of such operator 
with Dirichlet boundary conditions is $\det\mathcal{O}=\psi\left(b\right)$, where $\psi$ is the unique solution to the initial value problem to $\left(\mathcal{O}\psi\right)(x)=0$ 
with $\psi(a)=0$ and $\psi'(a)=1$. In fact, arguing that the product of increasing eigenvalues of the Sturm-Liouville problem is divergent, it rather yields an answer for the ratio 
of determinants of operators whose eigenvalues share the same asymptotic behavior~\cite{Dunne}. Consequently, the subsequent solutions, as explicit expressions for the pertinent functional determinants, 
should be regarded formal . In the remainder of the appendix the Gel'fand-Yaglom 
method is applied to the spectral problems defined by the operators in (\ref{molgaY}) and (\ref{Reference}) over the interval $[\epsilon,R]$.

\subsection{Functional determinants for the first surface}

We will evaluate the six functional determinants listed in (\ref{molgaY}). The solutions to the initial value problem to be derived here,
\be
-\psi''_{n}(\sigma) + V_{n}(\sigma)\psi_{n}(\sigma) = 0 \ , \quad \psi_{n}(\epsilon)=0 \ , \quad \psi'_{n}(\epsilon) = 1 \ ,
\ee
have been already found in~\cite{Kruczenski,Kim}. Recall that the symmetry $\widehat{Z}_{n}=\widehat{Z}_{-n}$ allows one to restrict the mode number $n$ to be a non-negative integer.  

\no
1.
If $n>1$, the solution to $\widehat{\D}_{B,1}\psi_{n}(\sigma) = 0$ with $\psi_{n}(\epsilon)=0$ and $\psi'_{n}(\epsilon)=1$ is
\ba
\label{Anterior}
\psi_{n}(\sigma) & \!\! = \!\! & \frac{1}{2n(n^2-1)} \big[ \left(n+\cotanh\epsilon\right)\left(n-\cotanh\sigma\right)\e^{n\left(\sigma-\epsilon\right)} \nonumber \\ 
& \!\! - \!\! & \left(n-\cotanh\epsilon\right)\left(n+\cotanh\sigma\right)\e^{-n\left(\sigma-\epsilon\right)} \big] \ ,
\ea
whose large $R$ limit is
\be
\psi_{n}(R) = \frac{\left(n+\cotanh\epsilon\right)}{2n\left(n+1\right)}\e^{n(R-\epsilon)} + \, \mathcal{O}\left(\e^{\left(n-2\right)R}\right) \ .
\ee
If $n=1$, the solution to $\widehat{\D}_{B,1}\psi_{1}(\sigma)=0$ with $\psi_{1}(\epsilon)=0$ and $\psi'_{1}(\epsilon)=1$ is
\be
\psi_{1}(\sigma) = \frac{\cosech\epsilon\cosech\sigma}{4} \big[ \sinh\left(2\sigma\right)-\sinh\left(2\epsilon\right)-2\left(\sigma-\epsilon\right) \big] \ ,
\ee
expression that may be obtained as a limit of (\ref{Anterior}). Its large $R$ asymptotic expansion is
\be
\psi_{1}(R) = \frac{\left(1+\cotanh\epsilon\right)}{4}\e^{R-\epsilon} + \, \mathcal{O}\left(R\e^{-R}\right) \ .
\ee
If $n=0$, the solution to $\widehat{\D}_{B,1}\psi_{0}(\sigma)=0$ with $\psi_{0}(\epsilon)=0$ and $\psi'_{0}(\epsilon)=1$ is
\be
\psi_{0}(\sigma) = \cotanh\sigma-\cotanh\epsilon + \cotanh\epsilon \left(\sigma-\epsilon\right) \ ,
\ee
which again may be obtained from (\ref{Anterior}) as a limit. Its large $R$ asymptotic expansion is
\be
\psi_{0}(R) = \cotanh\epsilon\,R + \, \mathcal{O}\left(1\right) \ .
\ee

\no
2.
If $n>0$, the solution to $\widehat{\D}_{B,2} \psi_{n}(\sigma) = 0$ with $\psi_{n}(\epsilon) = 0$ and $\psi'_{n}(\epsilon)=1$ is
\be
\label{Previo}
\psi_{n}(\sigma) = \frac{\sinh(n\left(\sigma-\epsilon\right))}{n} \ ,
\ee
whose large $R$ limit is
\be
\label{Desarrollo}
\psi_{n}(R) = \frac{\e^{n(R-\epsilon)}}{2n} + \, \mathcal{O}\left(\e^{-nR}\right) \ .
\ee
If $n=0$, the solution to $\widehat{\D}_{B,2}\psi_{0}(\sigma)=0$ with $\psi_{0}(\epsilon)=0$ and $\psi'_{0}(\epsilon)=1$ is
\be
\psi_{0}(\sigma) = \sigma-\epsilon \ .
\ee
that may be obtained as a limit of (\ref{Previo}). Its large $R$ asymptotic expansion is
\be
\psi_{0}(R) = R+ \, \mathcal{O}(1) \ .
\ee

\no
3.
If $n>1$, the solution to $\widehat{\D}_{F,1}^{+}\psi_{n}(\sigma)=0$ with $\psi_{n}(\epsilon)=0$ and $\psi'_{n}(\epsilon)=1$ is
\ba
\label{Precedente}
\psi_{n}(\sigma) & \!\! = \!\! & \frac{\sqrt{\cosech\epsilon\cosech\sigma}}{2n(n-1)} \big[\e^{\frac{1}{2}\left(\sigma+\epsilon\right)}(n-1)\sinh\left(n\left(\sigma-\epsilon\right)\right) \nonumber \\
& \!\! - \!\! & \e^{-\frac{1}{2}\left(\sigma+\epsilon\right)}n\sinh\left(\left(n-1\right)\left(\sigma-\epsilon\right)\right) \big] \ ,
\ea
whose large $R$ limit is
\be
\label{Uno}
\psi_{n}(R) = \frac{\e^{-\left(n-\frac{1}{2}\right)\epsilon}}{2n\sqrt{2\sinh\epsilon}}\e^{nR}+ \, \mathcal{O}\left(\e^{\left(n-2\right)R}\right) \ .
\ee
If $n=1$, the solution $\widehat{\D}_{F,1}^{+}\psi_{1}(\sigma)=0$ with $\psi_{1}(\epsilon)=0$ and $\psi'_{1}(\epsilon)=1$ is
\be
\psi_{1}(\sigma) = \frac{1}{2\sqrt{\sinh\epsilon\sinh\sigma}}\big[\e^{\frac{1}{2}\left(\sigma+\epsilon\right)}\sinh\left(\sigma-\epsilon\right)-\e^{-\frac{1}{2}\left(\sigma+\epsilon\right)}\left(\sigma-\epsilon\right)\big] \ ,
\ee
that may be derived as a limit of (\ref{Precedente}). Its large $R$ asymptotic expansion is
\be
\psi_{1}(R) = \frac{\e^{-\frac{1}{2}\epsilon}}{2\sqrt{2\sinh\epsilon}}\e^{R} + \, \mathcal{O}\left(R\e^{-R}\right) \ .
\ee
If $n=0$, the solution $\widehat{\D}_{F,1}^{+}\psi_{0}(\sigma)=0$ with $\psi_{0}(\epsilon)=0$ and $\psi'_{0}(\epsilon)=1$ is
\be
\label{Zero}
\psi_{0}(\sigma) = \frac{1}{2\sqrt{\sinh\epsilon\sinh\sigma}}\big[\e^{\frac{1}{2}\left(\sigma+\epsilon\right)}\left(\sigma-\epsilon\right)-\e^{-\frac{1}{2}\left(\sigma+\epsilon\right)}\sinh\left(\sigma-\epsilon\right)\big] \ ,
\ee
which again is derivable from (\ref{Precedente}) as a limit. Its large $R$ asymptotic expansion is
\be
\psi_{0}(R) = \frac{\e^{\frac{1}{2}\epsilon}}{\sqrt{2\sinh\epsilon}}R + \, \mathcal{O}\left(1\right) \ .
\ee

\no
4.
If $n>0$, the solution to $\widehat{\D}_{F,1}^{-}\psi_{n}(\sigma)=0$ with $\psi_{n}(\epsilon)=0$ and $\psi'_{n}(\epsilon)=1$ is
\ba
\psi_{n}(\sigma) & \!\! = \!\! & \frac{\sqrt{\cosech\epsilon\cosech\sigma}}{2n(n+1)} \big[ \e^{\frac{1}{2}\left(\sigma+\epsilon\right)}(n+1)\sinh\left(n\left(\sigma-\epsilon\right)\right) \nonumber \\
& \!\! - \!\! & \e^{-\frac{1}{2}\left(\sigma+\epsilon\right)}n\sinh\left(\left(n+1\right)\left(\sigma-\epsilon\right)\right) \big] \ ,
\ea
whose large $R$ limit is 
\be
\label{Dos}
\psi_{n}(R) = \frac{\big[(2n+1)\sinh\epsilon+\cosh\epsilon\big]\e^{-\left(n+\frac{1}{2}\right)\epsilon}}{2n(n+1)\sqrt{2\sinh\epsilon}}\e^{nR} + \, \mathcal{O}\left(\e^{\left(n-2\right)R}\right) \ .
\ee
If $n=0$, the solution to $\widehat{\D}_{F,1}^{-}\psi_{0}(\sigma)=0$ with $\psi_{0}(\epsilon)=0$ and $\psi'_{0}(\epsilon)=1$ is (\ref{Zero}).

\no
5.
If $n\ge0$, the solution to $\widehat{\D}_{F,2}^{+}\psi_{n}(\sigma)=0$ with $\psi_{n}(\epsilon)=0$ and $\psi'_{n}(\epsilon)=1$ is
\be
\label{noituloS}
\psi_{n}(\sigma) = \sqrt{\sinh\epsilon\sinh\sigma}\e^{\left(n-\frac{1}{2}\right)(\sigma+\epsilon)}\big[\B (\e^{-2\epsilon};n,0 ) - \B (\e^{-2\sigma};n,0) \big] \ ,
\ee
where $\B(x;a,b)$ is the incomplete Euler beta function. If $n>0$ is further satisfied, the large $R$ limit of (\ref{noituloS}) is
\be
\label{Tres}
\psi_{n}(R) = \sqrt{\frac{\sinh\epsilon}{2}} \B (\e^{-2\epsilon};n,0) \e^{\left(n-\frac{1}{2}\right)\epsilon}\e^{nR} + \, \mathcal{O}\left(\e^{\left(n-2\right)R}\right) \ .
\ee
On the contrary, if $n=0$, where the solution reduces to
\be
\label{oreZ}
\psi_{0}(\sigma) = \sqrt{\sinh\epsilon\sinh\sigma}\e^{-\frac{1}{2}\left(\sigma+\epsilon\right)}\left[2\left(\sigma-\epsilon\right)+\log\left(\frac{1-\e^{-2\sigma}}{1-\e^{-2\epsilon}}\right)\right] \ ,
\ee
its large $R$ limit is
\be
\label{Linear}
\psi_{0}(R) = \e^{-\frac{1}{2}\epsilon}\sqrt{2\sinh\epsilon}\,R + \, \mathcal{O}\left(1\right) \ .
\ee

\no
6.
If $n\ge0$, the solution to $\widehat{\D}_{F,2}^{-}\psi_{n}(\sigma)=0$ with $\psi_{n}(\epsilon)=0$ and $\psi'_{n}(\epsilon)=1$ is
\be
\label{Psi}
\psi_{n}(\sigma) = \sqrt{\sinh\epsilon\sinh\sigma}\e^{-\left(n+\frac{1}{2}\right)\left(\sigma+\epsilon\right)} \big[\B (\e^{-2\epsilon};-n,0) - \B (\e^{-2\sigma};-n,0) \big] \ .
\ee
If $n>0$ is fulfilled, the large $R$ limit of (\ref{Psi}) is
\be
\label{Cuatro}
\psi_{n}(R) = \sqrt{\frac{\sinh\epsilon}{2}}\frac{\e^{-\left(n+\frac{1}{2}\right)\epsilon}}{n}\e^{nR} + \, \mathcal{O}\left(\e^{\left(n-2\right)R}\right) \ ,
\ee
whereas it is given by (\ref{Linear}) if $n=0$, where the solution is again (\ref{oreZ}).

\subsection{Functional determinants for the second surface}

Consider now the functional determinants appearing in (\ref{Reference}). The solutions to the initial value problem to be derived here,
\be
-\psi''_{p}(\sigma) + V_{p}(\sigma)\psi_{p}(\sigma)=0 \ , \quad \psi_{p}(\epsilon)=0 \ , \quad \psi'_{p}(\epsilon)=1 \ ,
\ee
have been already reported in~\cite{Kruczenski}. Again, the symmetry $\widehat{Z}_{p}=\widehat{Z}_{-p}$ allows one to restrict $p$ to be non-negative without loss 
of generality.~\footnote{The case $p=0$ is excluded, as in~\cite{Kruczenski}, since the dependence on the regulator $R$ does not cancel when the solutions are combined in the proper ratio. It should correspond to a non-normalizable zero mode.}

\no
1.
The solution to $\widehat{\D}_{B,1}\psi_{p}(\sigma)=0$ with $\psi_{p}(\epsilon)=0$ and $\psi'_{p}(\epsilon)=1$ is
\be
\psi_{p}(\sigma) = \frac{1}{p^3\epsilon\sigma}\left[\left(p^2\epsilon\sigma-1\right)\sinh\left(p\left(R-\epsilon\right)\right)+p\left(\sigma-\epsilon\right)\cosh\left(p\left(\sigma-\epsilon\right)\right)\right] \ ,
\ee
whose large $R$ limit is 
\be
\psi_{p}\left(\sigma\right)=\frac{\left(\epsilon p+1\right)}{2 p^{2}\epsilon}\e^{p(R-\epsilon)} + \, \mathcal{O}\left(\frac{\e^{pR}}{R}\right) \ .
\ee

\no
2.
The solution to $\widehat{\D}_{B,2}\psi_{p}(\sigma)=0$ with $\psi_{p}(\epsilon)=0$ and $\psi'_{p}(\epsilon)=1$ is given by equation (\ref{Previo}) with $n$ replaced by $p$, 
and thus its large $R$ limit follows again from (\ref{Desarrollo}).

\no
3.
The solution to $\widehat{\D}_{F,1}^{+}\psi_{p}(\sigma)=0$ with $\psi_{p}(\epsilon)=0$ and $\psi'_{p}(\epsilon)=1$ is
\be
\psi_{p}\left(\sigma\right) = \frac{1}{4p^2\sqrt{\epsilon\sigma}}\left[\left(2p\sigma-1\right)\e^{p\left(\sigma-\epsilon\right)}-\left(2p\epsilon-1\right)\e^{-p\left(\sigma-\epsilon\right)}\right] \ ,
\ee
whose large $R$ limit is
\be
\psi_{p}\left(R\right) = \frac{1}{2p}\sqrt{\frac{R}{\epsilon}}\e^{p(R-\epsilon)} + \, \mathcal{O}\left(\frac{\e^{p R}}{\sqrt{R}}\right) \ .
\ee

\no
4.
The solution to $\widehat{\D}_{F,1}^{-}\psi_{p}(\sigma)=0$ with $\psi_{p}(\epsilon)=0$ and $\psi'_{p}(\epsilon)=1$ is
\be
\psi_{p}\left(\sigma\right)=\frac{1}{4p^2\sqrt{\epsilon\sigma}}\left[\left(2p\epsilon+1\right)\e^{p\left(\sigma-\epsilon\right)}-\left(2p\sigma+1\right)\e^{-p\left(\sigma-\epsilon\right)}\right] \ ,
\ee
whose large $R$ limit is
\be
\psi_{p}\left(R\right) = \frac{\left(2p\epsilon+1\right)}{4p^2\sqrt{\epsilon R}}\e^{p(R-\epsilon)} +  \, \mathcal{O}\left(\e^{-p R}\right) \ .
\ee

\no
5.
The solution to $\widehat{\D}_{F,2}^{+}\psi_{p}(\sigma)=0$ with $\psi_{p}(\epsilon)=0$ and $\psi'_{p}(\epsilon)=1$ is
\be
\psi_{p}\left(\sigma\right) = \e^{p\left(\epsilon+\sigma\right)}\sqrt{\epsilon\sigma} \big[ \Ei_{1}\left(2p\epsilon\right)-\Ei_{1}\left(2p\sigma\right) \big] \ ,
\ee
where $\Ei_{1}(x)$ is the exponential integral. Its large $R$ limit is
\be
\psi_{p}\left(R\right) = \sqrt{\epsilon R}\Ei_{1}\left(2p\epsilon\right) \e^{p(R+\epsilon)} + \, \mathcal{O}\left(\frac{\e^{p R}}{\sqrt{R}}\right) \ .
\ee

\no
6.
The solution to $\widehat{\D}_{F,2}^{-}\psi_{p}(\sigma)=0$ with $\psi_{p}(\epsilon)=0$ and $\psi'_{p}(\epsilon)=1$ is
\be
\psi_{p}\left(\sigma\right) = \e^{-p\left(\epsilon+\sigma\right)}\sqrt{\epsilon\sigma}\big[\Ei\left(2p\sigma\right)-\Ei\left(2p\epsilon\right)\big] \ ,
\ee
with $\Ei(x)$ the analytic continuation of $\Ei_{1}(x)$, which satisfies $\Ei(x)=-\Ei_{1}\left(-x\right)$ when $x<0$. In the limit of large $R$ it reduces to
\be
\psi_{p}\left(R\right) = \sqrt{\frac{\epsilon}{R}}\frac{1}{2 p}\e^{p(R-\epsilon)} + \, \mathcal{O}\left(\frac{\e^{p R}}{R^{\frac{3}{2}}}\right) \ .
\ee

%%%%%%%%%%%%%%%%%%%%%%%%%%%%%%%%%%%%%%%%%%%%%%%%%%%%%%%%%%%%%%%%%%
%%%%%%%%%%%%%%%%%%%%%%%%%%%%%%%%%%%%%%%%%%%%%%%%%%%%%%%%%%%%%%%%%%

\section{The Abel-Plana formula}

\label{Plana}

Let $f(z)$ be certain holomorphic function over the upper complex plane decaying rapidly enough at infinity. For such a function the Abel-Plana formula states that
\be
\label{Abel}
\overset{\infty}{\underset{n=1}{\sum}}f(n) = - \frac{1}{2}f(0) + \int_{0}^{\infty}\dif x f(x)+i \int_{0}^{\infty}\dif x \, \frac{f(ix)-f(-ix)}{\e^{2\pi x}-1} \ .
\ee
The formula cannot be rigorously employed for the sums considered in this article as the functions of interest do not satisfy the proper requirements. 
Nonetheless, it allows us to formally achieve the integral expressions in which the discussion of the main text is based. In fact, we will prove that the second integral 
in the Abel-Plana formula is convergent for all the cases under study.  We will assume that the infrared regulator is positive as in the main text. Besides, we must stress again 
that the subsequent divergent integrals here are to be regarded formally. 

The sums to be performed may be split into three types. We will consider first the linear contribution, where $f(x)=(2x-1)\epsilon$. By means of (\ref{Abel}), we are led to
\be
\label{Lineal}
\overset{\infty}{\underset{n=1}{\sum}}\left(2x-1\right)\epsilon=\frac{1}{2}\epsilon + \!  \int_{0}^{\infty}\dif x (2x-1)\epsilon-4 \!  \int_{0}^{\infty}\dif x \frac{x\epsilon}{\e^{2\pi x}-1} 
=\frac{\epsilon}{3} + \!  \int_{0}^{\infty}\dif x (2x-1)\epsilon \ .
\ee
Next, we will move to the linear logarithmic contribution, where $f(x)=\log\left(ax+b\right)$ and $a,b>0$. By means of (\ref{Abel}), we are led to
\be
\overset{\infty}{\underset{n=1}{\sum}}\log\left(an+b\right)=-\frac{1}{2}\log b + \int_{0}^{\infty}\dif x \log(ax+b)-2\int_{0}^{\infty}\dif x \frac{\arctan\left(ax/b\right) }{\e^{2\pi x}-1} \ .
\ee
The last integral can be performed explicitly,
\be
\label{Gradshteyn}
\int_{0}^{\infty}\dif x \frac{\arctan\left(ax/b\right) }{\e^{2\pi x}-1}=\frac{1}{2}\left[\log\Gamma\left(\frac{b}{a}\right)-\left(\frac{b}{a}-\frac{1}{2}\right)\log\frac{b}{a}+\frac{b}{a}-\frac{1}{2}\log\left(2\pi\right)\right] ,
\ee
that vanishes when $a\rightarrow 0$. If we specify the linear logarithms involved in equation (\ref{amuS}) above, we find that
\be
\label{Logaritmo}
\begin{split}
&\overset{\infty}{\underset{n=1}{\sum}}\log\left(n+1\right)=\int_{0}^{\infty}\dif x \log(x+1)-2\int_{0}^{\infty}\dif x \frac{\arctan x }{\e^{2\pi x}-1} \ , \\
&\overset{\infty}{\underset{n=1}{\sum}}\log\left(n+\cotanh\epsilon\right)=\frac{1}{2}\log (\tanh\epsilon) + \! \int_{0}^{\infty}\dif x \log(x+\cotanh\epsilon) \\
& -2 \! \int_{0}^{\infty}\dif x \frac{\arctan\left(x\tanh\epsilon\right) }{\e^{2\pi x}-1} \ , \\
&\overset{\infty}{\underset{n=1}{\sum}}\log\left(\left(2n+1\right)\sinh\epsilon+\cosh\epsilon\right)=-\epsilon+\int_{0}^{\infty}\dif x \log\left(\left(2x+1\right)\sinh\epsilon+\cosh\epsilon\right) \\
&-2\int_{0}^{\infty}\dif x \frac{\arctan\left(2x\tanh\epsilon/\left(1+\tanh\epsilon\right)\right) }{\e^{2\pi x}-1} \ .
\end{split}
\ee
where the finite integrals can be calculated using expression (\ref{Gradshteyn}). Finally, we will consider the logarithmic contribution of the Euler beta function, 
where $f(x)=\log\B(\e^{-2\epsilon};x,0)$. This case requires us to single out the first contribution and shift by a unity the index of the sum, on account of the strict divergence 
of the Euler beta function at $n=0$. If we proceed in this way, we conclude that
\ba
\label{Beta}
\overset{\infty}{\underset{n=1}{\sum}}\log\B(\e^{-2\epsilon};n,0)& \!\! = \!\! & \frac{1}{2}\log\B(\e^{-2\epsilon};1,0)+\int_{0}^{\infty}\dif x \log\B(\e^{-2\epsilon};x+1,0) \nonumber \\
& \!\! + \!\! & \int_{0}^{\infty}\dif x \frac{2\arctan\left(g_{1}(x)/g_{2}(x)\right)}{\e^{2\pi x}-1} \ ,
\ea
with $g_{1}(x)$ and $g_{2}(x)$ given by
\be
g_{1}(x) = \overset{\infty}{\underset{k=1}{\sum}}\frac{x\e^{-2k\epsilon}}{(k+1)^2+x^2} \ , \quad g_{2}(x)=\overset{\infty}{\underset{k=1}{\sum}}\frac{(k+1)\e^{-2k\epsilon}}{(k+1)^2+x^2} \ ,
\ee
where we have made use of the series representation
\be
\B(\e^{-2\epsilon};x,0) = \overset{\infty}{\underset{k=0}{\sum}}\frac{\e^{-2\left(k+x\right)\epsilon}}{k+x} \ ,
\ee
valid if $\epsilon>0$ and $x>0$, which is the case above. The integrand of the last term has a removable singularity at $x=0$ and thus the contribution of its associated integral is finite. 

%%%%%%%%%%%%%%%%%%%%%%%%%%%%%%%%%%%%%%%%%%%%%%%%%%%%%%%%%%%%%%%%%%
%%%%%%%%%%%%%%%%%%%%%%%%%%%%%%%%%%%%%%%%%%%%%%%%%%%%%%%%%%%%%%%%%%

\end{document}